\documentclass[aps,pra,twocolumn,showpacs,superscriptaddress,groupedaddress]{revtex4}  % for review and submission
\usepackage{graphicx}  % needed for figures
\usepackage{dcolumn}   % needed for some tables
\usepackage{bm}        % for math
\usepackage{amssymb}   % for math
\usepackage{braket}
\usepackage{gensymb}
\usepackage{amssymb,amsmath,amsbsy}
\newcommand{\persec}{\textrm{s}^{-1}}

\begin{document}

\title{Coherent frequency up-conversion of microwaves to the optical telecommunications band in an Er:YSO crystal}

\author{Xavier Fernandez-Gonzalvo}
\author{Yu-Hui Chen}
\affiliation{The Dodd-Walls Centre for Photonic and Quantum Technologies \& Department of Physics, University of Otago, 730 Cumberland Street, Dunedin, New Zealand.}
\author{Chunming Yin}
\author{Sven Rogge}
\affiliation{Centre of Excellence for Quantum Computation and Communication Technology, School of Physics, University of New South Wales, Sydney, New South Wales 2052, Australia}
\author{Jevon J. Longdell}
\affiliation{The Dodd-Walls Centre for Photonic and Quantum Technologies \& Department of Physics, University of Otago, 730 Cumberland Street, Dunedin, New Zealand.}

\email{jevon.longdell@otago.ac.nz}

\date{\today}

\begin{abstract}
The ability to convert quantum states from microwave photons to optical photons is important for hybrid system approaches to quantum information processing. In this paper we report the up-conversion of a microwave signal into the optical telecommunications wavelength band using erbium dopants in a yttrium orthosilicate crystal via stimulated Raman scattering. The microwaves were applied to the sample using a 3D copper loop-gap resonator and the coupling and signal optical fields were single passed. The conversion efficiency was low, in agreement with a theoretical analysis, but can be significantly enhanced with an optical resonator.
\end{abstract}


\maketitle

\section{Introduction}

Superconducting qubits are a rapidly advancing part of quantum information science. The ability to reach deep into the strong coupling regime of cavity QED using microwaves has revolutionized quantum optics in the microwave regime \cite{kubo2010,schu2010,teuf2011,prob2013}, and allows the coupling between superconducting qubits and a broad range of microwave frequency quantum systems \cite{laha2009,zhu2011}. Distribution and storage of microwave quantum states, however, present difficult challenges. A way around this problem would be to convert quantum states of microwave photons into optical photons and vice versa. This would allow long distance propagation of quantum states between superconducting qubit nodes using optical fibres, and it would also allow for quantum memories for light to be used \cite{long2005,hoss2009,usma2010,hedg2010,timo2013}, which are currently more developed than their microwave counterparts \cite{clel2004,sill2007,wu2010,timo2011}. Quantum  frequency conversion has been achieved between optical frequencies \cite{tanz2005,mcgu2010,radn2010,zask2012,albr2014}, and recently between microwave frequencies \cite{inom2014}. However, so far, quantum frequency conversion from the microwave to the optical domain remains an unsolved challenge.

There are a number of approaches being investigated for the up-conversion process. Opto-mechanical approaches \cite{amir2011,hill2012,mcge2013,rega2011,andr2014} currently have the highest reported efficiencies and can achieve MHz bandwidths. In such approaches both an optical and a microwave resonators are parametrically coupled through a micro-mechanical resonator. In order to have quiet frequency conversion this rather low frequency intermediate mechanical resonator needs to be cooled to its quantum ground state, and this is currently challenging. Another approach is to use conventional non-linear optical materials to make resonantly enhanced modulators \cite{ilch2002,stre2009,tsan2010}.

Two recent proposals \cite{will2014,obri2014} have suggested using rare earth doped solids, with a particular focus on erbium doped yttrium orthosilicate (Er:YSO). Er:YSO has many attractive features for frequency up-conversion: it has narrow inhomogeneous and homogeneous linewidths for its ${^4I_{15/2}} \leftrightarrow {^4I_{13/2}}$ optical transition \cite{bott2006}, and the wavelength of this transition is in the telecommunications band, where propagation losses in optical fibres are minimized. Because Er${^{3+}}$ is a Kramer's ion (odd number of $4f$ electrons), for the nuclear spin free isotopes (all but $^{167}$Er), the ground state is doubly degenerate. It also has rather large effective $g$ values \cite{sun2008,guil2006}, such that microwave frequency splittings can be achieved with only modest magnetic fields.

\begin{figure}[b]
\includegraphics[scale=0.27]{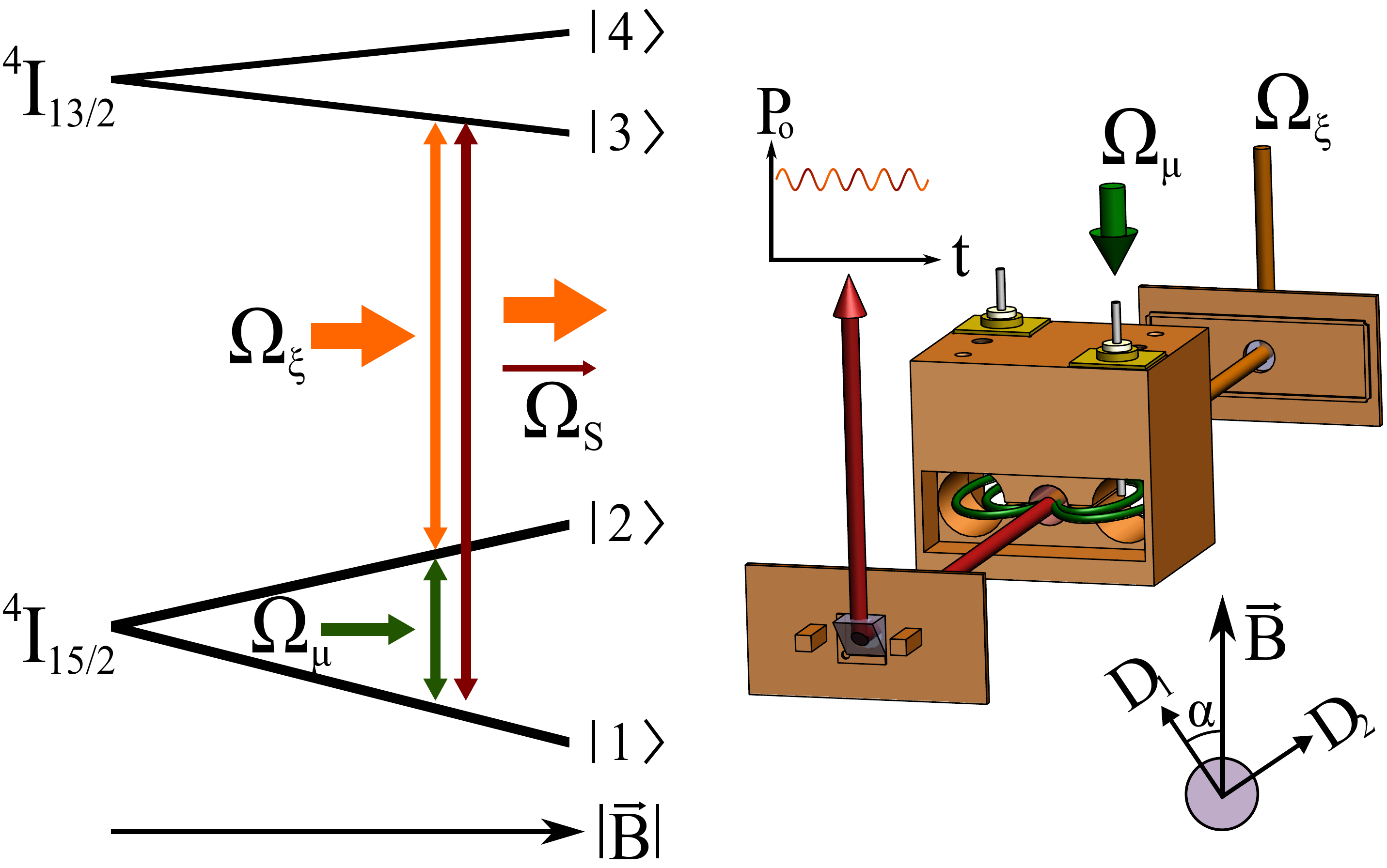}
\caption{\label{fig:levelssetup} (Color online) \emph{Left:} the ${^4I_{15/2}}$ and ${^4I_{13/2}}$ levels in Er:YSO are Zeeman split under the presence of an external magnetic field $\vec{B}$. The Raman heterodyne signal is produced when a microwave field $\Omega_\mu$ and an optical field $\Omega_\xi$ drive two transitions in a three level atom. A coherence is produced on the third transition which generates an optical signal field $\Omega_S$. This can be detected as a beat note on the optical drive field, i.e. a modulation in the optical output power $P_o$ at the same frequency as the microwave field. \emph{Right:} depiction of the experimental setup. A copper made loop-gap resonator holds an Er:YSO sample inside. Light is coupled in and out using a pair of prisms. Microwaves are coupled via two straight antennas. The magnetic field is applied in the vertical direction, parallel to the $D_1$-$D_2$ plane of the crystal, at an angle $\alpha$ as measured from $D_1$.}
\end{figure}

In the present letter we report the up-conversion of a microwave signal into the optical telecommunications band using rare earth ion dopants in a crystal, by performing microwave Raman heterodyne spectroscopy in Er:YSO. Raman heterodyne spectroscopy with radio frequency (ca. 0-200\,MHz) is a commonly used technique for nuclear spins in rare earth dopants \cite{mlyn1983,long2002,lovr2011,eric1987,long2006}. It has also been demonstrated in the microwave regime in ruby \cite{bing1998,schw2002} and metalloprotiens \cite{bing2000}. These systems, however, are not as attractive for the realization of quantum frequency conversion because they exhibit much broader optical lines.

\section{Frequency up-conversion}

Raman heterodyne spectroscopy uses the three wave mixing that occurs from three energy levels in a  $\Delta$ configuration, as shown in Fig.~\ref{fig:levelssetup}. To enhance the efficiency of the process we use a microwave resonator for the lowest frequency field. The $^4I_{15/2}$ ground state of Er:YSO is Zeeman split under the presence of an applied magnetic field $\vec{B}$, making the ${\Ket{1}\leftrightarrow\Ket{2}}$ transition resonant with the microwave cavity. When the input microwave field $\Omega_{\mu}$ is applied it generates a coherence between levels $\Ket{1}$ and $\Ket{2}$. Simultaneously, the optical coupling field $\Omega_{\xi}$ drives a second coherence between levels $\Ket{2}$ and $\Ket{3}$. The presence of these two coherences generates a third one between levels $\Ket{1}$ and $\Ket{3}$, which gives an output signal field $\Omega_S$ at a frequency equal to the sum of the frequencies of the microwave and the coupling fields. As long as the sample is small compared to the wavelength of the microwave field the signal field will be generated in the same spatial mode as the coupling beam. The signal field can then be readily detected in a photodiode as a heterodyne beat note on the coupling beam.

\subsection{Experimental realisation}

The crystalline structure of YSO belongs to the $C^6_{2h}$ symmetry group, with two crystalographically inequivalent sites where erbium can replace yttrium. In this work we focused on `Site~1' with a transition wavelength of $\lambda_1$~=~1536.478\,nm \cite{li1992}. YSO has three orthogonal optical extinction axes $D_1$, $D_2$ and $b$. We use a cylindrical Er:YSO sample of 4.95\,mm diameter by 12\,mm length, with an erbium number concentration of 0.001\%. The optical $b$ axis of the crystal is aligned along the length of the cylinder, and so the $D_1$-$D_2$ plane is parallel to the end faces. The sample sits inside a copper three-dimensional loop-gap microwave resonator, with a resonant frequency of 4.9\,GHz and a linewidth of 16\,MHz (quality factor $Q\simeq$\,300). This kind of resonator provides very good filling factors ($\sim$0.8) and makes optical coupling to the sample a simple task, since two optical windows can be opened in the end caps at a null point of the surface currents, thus not affecting the properties of the cavity very much. Input and output microwave powers are coupled with a pair of straight antennas inside the cavity space. The input light, at 1536\,nm, is coupled into and out of the sample with the aid of a pair of coupling prisms, and fibre coupled collimators. The input fibre is a single mode fibre, while for the output one we use a multi-mode fibre for ease of coupling. A superconducting magnet generates a magnetic field perpendicular to the longitudinal direction of the sample (i.e. in the $D1$-$D2$ plane) between 0 and 300\,mT. The angle $\alpha$, measured from $D_1$ to $\vec{B}$, can be varied by rotating the sample. A more detailed explanation of the complete experimental setup can be found in Appendix~\ref{App:setup}.

The strength for each of the optical transitions in Fig.~\ref{fig:levelssetup} is given by the product of the electronic transition dipole moment and the overlap of the two spin states.  This overlap is calculated by diagonalizing the spin Hamiltonian \cite{sun2008} an taking the inner product of the respective eigenstates. The orientation of the magnetic field has to be chosen carefully, so as to maximize the difference between the quantisation axes for the ground and excited states and thus allow $\Delta$ transitions. For the situation in which $\vec{B}$ is contained in the $D_1$-$D_2$ plane the calculated angle that maximizes the overlap between states $\Ket{2}$ and $\Ket{3}$ is  $\alpha_M = 29$\degree.

After the frequency conversion process, the AC component of the heterodyne signal detected in the photodiode is high-pass filtered and amplified, and sent into an RF spectrum analyser.  An inconvenient consequence of using a multi-mode fibre for the output light is that there is loss in the modulation due to dephasing of the different propagation modes. In our setup this loss is typically from 3 to 10\,dB, and it depends on the arrangement of the fibre. From the power detected by the spectrum analyser we can work out the optical power of the generated signal sideband $P_S$.

\subsection{Raman heterodyne spectroscopy}

\begin{figure*}
\includegraphics[scale=0.75]{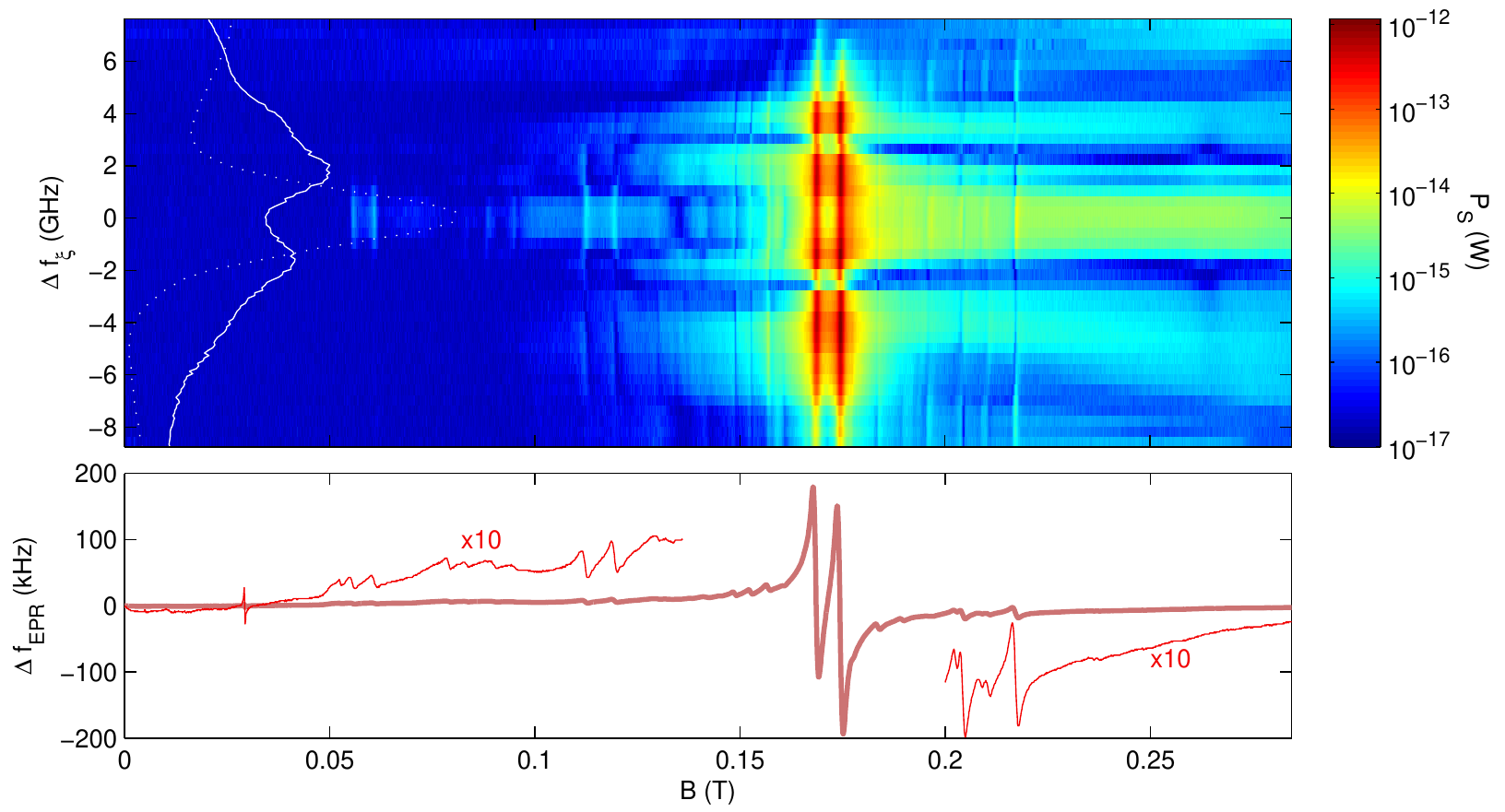}
\caption{\label{fig:raman} (Color online) \emph{Top}: Raman heterodyne spectroscopy on Er:YSO, showing frequency conversion from microwave to optical telecom frequencies. The strength of the magnetic field is plotted in the horizontal axis, and the coupling laser detuning is plotted on the vertical axis. The color scale indicates the power of the output signal field. On the left, the white dotted line represents the optical absorption spectrum for $\lvert\vec{B}\rvert$~=~0. The solid white line corresponds to the optical absorption spectrum for $\lvert\vec{B}\rvert$~=~0.178\,T. \emph{Bottom}: EPR spectrum of the Er:YSO sample. The regions away from the main peaks have been magnified for clarity. In both plots the presence of double peaks along the horizontal axis is due to misalignment in the magnetic field, breaking the magnetic degeneracy of the two inequivalent orientations of Er$^{+3}$ in YSO.}
\end{figure*}

Our Raman heterodyne spectroscopy results are presented in the color plot of Fig.~\ref{fig:raman}. The power of the generated signal field is measured as we scan the magnetic field and the coupling laser frequency $f_\xi$. On the left side, in white, we plot an optical absorption spectrum for $\lvert\vec{B}\rvert$~=~0 (dotted line)  and for $\lvert\vec{B}\rvert$~=~0.178\,T (solid line). Note that due to various etalon effects the background of these measurements is not constant. In the optical absorption spectrum for $\lvert\vec{B}\rvert$~=~0.178\,T the four optical transitions can be observed. The strong ones ($\Ket{1}\leftrightarrow\Ket{3}$ and $\Ket{2}\leftrightarrow\Ket{4}$) appear as peaks around $\Delta f_\xi=\pm$1.6\,GHz, while the weak ones ($\Ket{1}\leftrightarrow\Ket{4}$ and $\Ket{2}\leftrightarrow\Ket{3}$) appear as smaller shoulders at about $\Delta f_\xi=\pm$3.4\,GHz. The ratio between the absorption level of the weak and the strong lines is close to the expected value for $\alpha \simeq \alpha_M$. From the absorption measurements we can also extract an inhomogeneous broadening of the optical transition of $\sim$2.5\,GHz FWHM. Comparing the Raman heterodyne spectroscopy data with the absorption spectrum at $\lvert\vec{B}\rvert$~=~0.178\,T we see that the main four peaks in the color plot (in red) coincide with the absorption on each of the lines, as it is to be expected. It can also be seen that the peak signal is slightly higher for the lowest frequency peaks, which can be explained using hole burning arguments.

\subsection{Electron paramagnetic resonance}

Beneath the Raman heterodyne spectroscopy data is the electron paramagnetic resonance (EPR) spectrum of our sample. To take these EPR measurements we apply a frequency modulated (FM)  microwave signal into the input port of the microwave cavity and monitor the transmitted intensity using a lock-in amplifier. In this way we are able monitor the resonant frequency shift of our cavity ($\Delta f_{EPR}$) as is done in Pound frequency locking \cite{poun1946}. As the spin transitions are swept through resonance with the cavity they pull the resonator frequency first one way then the other, resulting in dispersive shaped peaks. The collection of vertical lines in the Raman heterodyne spectrum and the smaller peaks in the EPR spectrum are due to the $^{167}$Er isotope, which has non-zero nuclear spin ($I = 7/2$) and therefore exhibits hyperfine splitting even for $\lvert\vec{B}\rvert = 0$.

The EPR data presented in Fig.~\ref{fig:raman} shows a maximum frequency shift of around 180\,kHz. The measurements shown in the figure are taken for an input microwave power of 0\,dBm, which is enough to start to saturate the microwave transition in the absence of the optical field. For saturation-free measurements, at lower microwave powers, we get a maximum shift of around 260\,kHz, which agrees with our numerical simulations. From this EPR shift we can extract a cavity-atoms cooperativity factor of the order of $6\times10^{-2}$.

Comparing the Raman heterodyne and the EPR spectra we can see that most of the features present in the Raman heterodyne spectroscopy data are also replicated in the EPR data. The EPR peak present at $B\approx0.03$\,T we assign to the Er atoms in Site 2 of YSO. Because the optical transition for these atoms is at a different wavelength we don't see a signal in Raman heterodyne spectroscopy.

\section{Characterization of the conversion process}

In this section we characterise the conversion process by examining the dependency of the output signal power with the input microwave and coupling powers. We compare these measurements with a numerical model of our experiment, which we also use to find out the various losses in our setup. Finally we estimate the efficiency of the conversion process and discuss several ways by which it can be increased.

\subsection{Scaling with the input powers}

\begin{figure}
\includegraphics[scale=0.65]{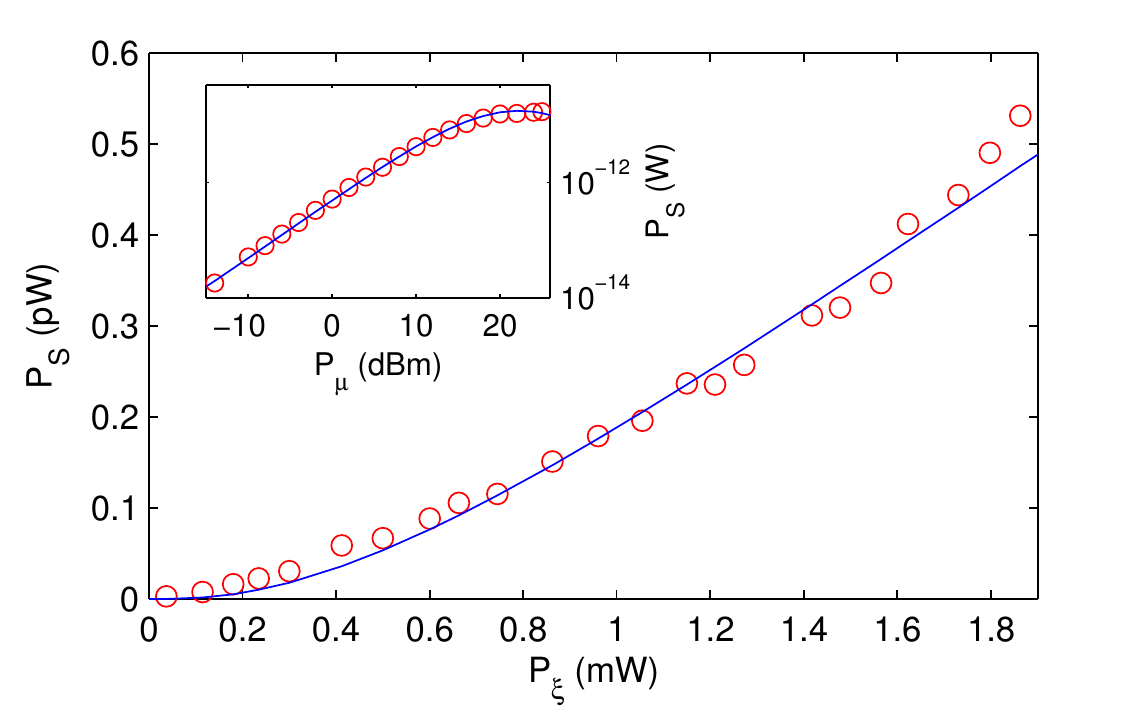}
\caption{\label{fig:psplasprf} (Color online) Signal power as a function of detected coupling laser power (main figure) and input microwave power (inset) in red, along with the corresponding theoretical predictions in blue. The faster-than-linear growth of $P_S$ versus $P_\xi$ shows optical cooling of the spins via optical pumping.}
\end{figure}

Figure~\ref{fig:psplasprf} shows, in red, the dependence of the signal field power with the input microwave power $P_\mu$ and the detected coupling laser power $P_\xi$. The laser coupling power is measured at the output of the system, and is not corrected for optical losses between the sample and the power meter. In blue we plot the expectations for these measurements based on our theoretical model, briefly discussed below. The dependency of $P_S$ with $P_\mu$ follows the expected pattern for a three wave mixing process: it increases linearly until it reaches a saturation point, in our case around $P_\mu=$\,20\,dBm. These measurements are taken for $P_\xi=$\,1.8\,mW. 

The dependency of $P_S$ with $P_\xi$, however, doesn't follow a quadratic relation for large laser powers. The faster than quadratic rate at which the signal increases with the pump laser power is due to optical pumping improving the population difference between the two I$_{15/2}$ sublevels, lowering the effective spin temperature. This fact is particularly convenient since, for a low noise conversion process, the spins temperature will need to be small compared with the frequency of the input microwave field. These measurements are taken for an input microwave power $P_\mu=$\,0\,dBm.

\subsection{Numerical model and propagation losses}

To model the experiment and plot the blue lines in Fig.~\ref{fig:psplasprf}, we treat each erbium atom as a three level system and use standard master equation techniques. This is described in detail in Appendix~\ref{App:model}. The optical and spin dephasing times are not know precisely and are allowed to vary, as is the spin lifetime. In the fitting process we also introduce two free parameters $\zeta_\mu$ and $\zeta^{-1}_\xi$ which take into account the propagation losses of $P_\mu$ from the setup input to the microwave cavity and the inverse loss of $P_\xi$ from the photodiode detector to the sample. The fitted values for these loss and inverse loss parameters are $\zeta_\mu=$\,13.1\,dB and $\zeta^{-1}_\xi=$\,-6.4\,dB. It is hard to compare these two numbers with the measured losses of the setup, since losses are hard to quantify at low temperatures for practical reasons. The measured $\zeta_\mu$ at room temperature is about 8\,dB. In the regime where $P_S$ is proportional to $P_\mu$ the loss of heterodyne signal in the multi-mode fibre is equivalent to a loss in microwave power. This multi-mode fibre loss is measured to be between 3 and 10\,dB depending on the geometrical arrangement of the fibre. To these two numbers we have to add the effects of lowering the temperature in the coaxial microwave wires, which are unknown but can be expected to be on the order of a few dB. For the optical losses we can measure the loss from the sample to the detector at room temperature, and it is around 5\,dB. It is hard to make an estimation of how lowering the temperature will modify this number. We could observe a total decrease in transmitted power of -10\,dB through our complete optical setup after the system was cold, but this number takes into account both, input and output losses, which we can not measure separately with the system inside the cryostat. All in all, we consider our fitted loss parameters to be in reasonable agreement with our observations.

\subsection{Conversion efficiency}

By comparing the input and the signal field powers we can calculate a number conversion efficiency ${\eta_n = \frac{P_S}{P_\mu}\cdot\frac{f_\mu}{f_\xi}}$, where $f_\mu$ is the input microwave frequency. This efficiency $\eta_n$ accounts for the fraction of microwave photons converted into optical telecom photons. For a coupling power of $\sim$2\,mW and making the appropriate corrections for $\zeta_\mu$ and $\zeta^{-1}_\xi$ we get a conversion efficiency of $\mathcal{O}\left(10^{-12}\right)$. In order to get closer to the target of unity conversion efficiency the most important improvement will be to add a doubly resonant optical cavity (for the coupling and signal fields), which will improve the efficiency by a factor proportional to the finesse of the cavity squared $F^2$, where $F$ can be as high as $\mathcal{O}\left(10^5\right)$. On top of this effect, cavity enhancement of the coupling field should increase the effectiveness of the optical cooling of the spins, additionally increasing the efficiency of the conversion process. There are also numerous other improvements that can be made. A more homogeneous magnetic field is very desirable, since it would reduce the microwave inhomogeneous linewidth. The optical depth used in this experiment is also rather low (0.02\,mm$^{-1}$). Much larger optical depths have been observed in Er:YSO without the penalty of broader inhomogeneous lines \cite{dajc2014}. Astonishingly narrow absorption lines with good optical depth have also been reported for Erbium dopants in isotopically pure yttrium lithium fluoride \cite{chuk2000}. The Q-factor of $\sim$300 for our microwave resonator is also rather modest -- much higher Q-factors for copper resonators have been reported \cite{prob2014}

\section{Conclusions}

In summary, we have presented a novel way to convert microwave signals into the optical telecommunications band, by means of a cryogenically cooled rare earth sample inside a three-dimensional microwave cavity. We have matched our Raman heterodyne spectroscopy experimental results with a theoretical counterpart. Finally, although the efficiency of this initial demonstration is low, there are many ways to improve it, the most significant of which is enhancing the effect of the two optical fields with an optical resonator. Among possible designs for this optical resonator is the Fabry-Perot resonator as proposed in \cite{will2014} or a whispering gallery mode type resonator as investigated in \cite{stre2009}.

\section*{ACKNOWLEDGEMENTS}

We would like to acknowledge the Marsden Fund (Contract No. UOO1221) of the Royal Society of New Zealand  and the ARC Centre of Excellence for Quantum Computation and Communication Technology (CE110001027) for their support. S.R. acknowledges a Future Fellowship (FT100100589).

\appendix

\renewcommand\thefigure{\thesection.\arabic{figure}} 

\section{A note on coherence}\label{App:coherence}
\setcounter{figure}{0}
\setcounter{equation}{0}

While we haven't performed any direct measurement of phase preservation in the up-conversion process we can be certain that this process is a coherent one. The observed heterodyne peaks are a few tens of Hz wide. Both the optical and the microwave transitions in Er:YSO are much wider than that, hence the only explanation is that this signal is indeed generated in a Raman scattering process, which is inherently coherent. In other words: the only process in our sample that can generate such a narrow signal is a coherent process. In addition, we are performing heterodyne detection, so our measurements are only sensitive to that light which is coherent with the pump beam.

\begin{figure}[b]
\centering
\includegraphics[scale=0.7]{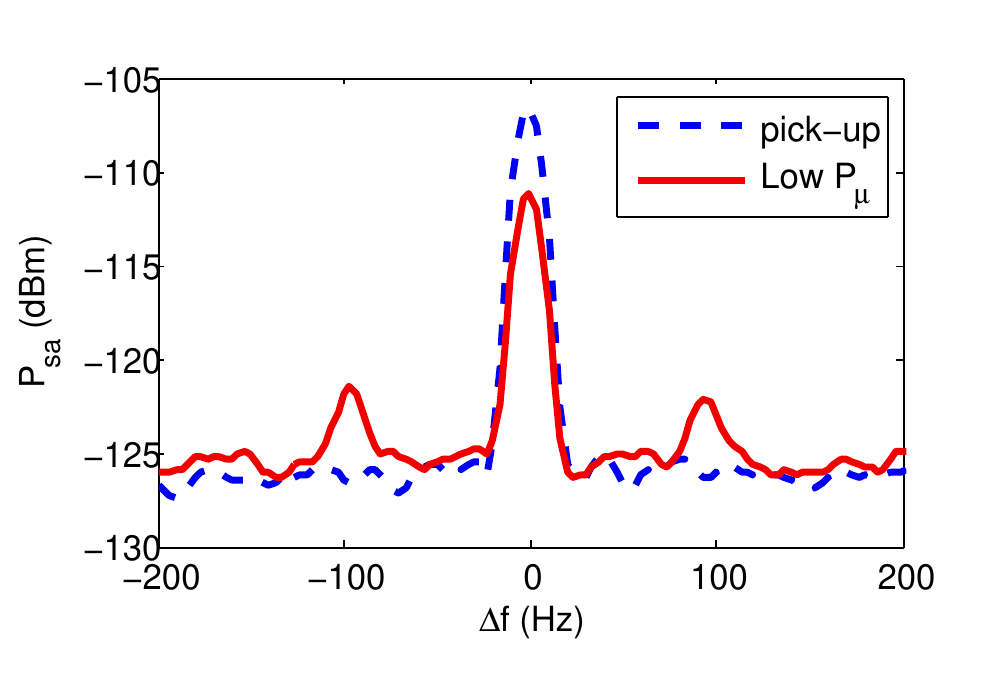}
\caption{\label{fig:pickupnoise} (Color online) Heterodyne signal as measured by the spectrum analyser, showing interference between the generated signal and the pick-up noise from the signal generator. The sidebands are attributed to sidebands in the laser frequency due to mechanical vibrations in the laser box.}
\end{figure}

To further support this claim there is the following experimental observation, represented in Fig.~\ref{fig:pickupnoise}. When measuring the Raman heterodyne signal with the spectrum analyser the noise background isn't flat. Instead we can observe a small peak at the signal generator's output frequency. This is pick-up noise coming directly from the signal generator's oscillator to the analyser (and the amplifier preceding it), and it is present even when the output of the signal generator is shut down (but the oscillator inside the signal generator is still on) or the optical detector is blocked. When measuring the heterodyne signal power versus the input microwave power, for high microwave powers the heterodyne signal is much bigger than the pick-up noise, and the later one can be neglected. For low microwave powers, however, the detected peak becomes smaller than the pick-up noise alone. This means that the detected signal and the pick-up noise interfere destructively, which in turn means they are coherent with each other. It is safe to assume that the pick-up noise will be coherent with the signal coming out of the signal generator since they come from the same source, so we can conclude that the converted signal is coherent with the input microwave signal, and therefore so is the conversion process.

\section{Theoretical model}\label{App:model}
\setcounter{figure}{0}
\setcounter{equation}{0}

To model the experiment and plot the blue lines in Fig.~3, we treat each erbium atom as a three level atom driven by a microwave field which connects the two lowest states $\ket{1}$ and $\ket{2}$, and an optical pump field which connects states $\ket{2}$ and $\ket{3}$. In the interaction picture we have the following Hamiltonian for a single atom:
\begin{equation} 
H = \delta_2\sigma_{22}+\delta_3\sigma_{33}+\Omega_\mu(\vec{r}\,)(\sigma_{12}+\sigma_{21})+\Omega_\xi(\vec{r}\,)(\sigma_{23}+\sigma_{32})~, 
\end{equation}
where $\delta_2$ is the detuning from the microwave cavity frequency, $\delta_3$ is the detuning from the coupling laser frequency and $\sigma_{ij}=\ket{\,i\,}\bra{\,j\,}$. Because in the center of the loop-gap resonator the magnetic field is rather uniform we can take the microwave Rabi frequency to be a constant ${\Omega_\mu (\vec{r}\,)\Rightarrow\Omega_\mu}$. We take the optical pump field as a plane wave propagating along the $z$ axis so the resulting Rabi frequency, which satisfies ${\Omega_\xi (\vec{r}\,) \equiv \Omega_\xi \cdot e^{ink_{32}z}}$, is represented by a travelling wave along the $z$ axis, where $k_{32}$ is the wave vector for light resonant with the $\ket{2} \leftrightarrow \ket{3}$ transition.

The dynamics of each of the atoms is governed by the master equation
\begin{equation}
\dot\rho = -i[\rho,H]+\mathcal{L}_\textrm{loss}\rho\label{eq:master}~,
\end{equation}
where $\rho$ is the density operator of a single atom and $\mathcal{L}_\textrm{loss}$ is the loss Lindblad superoperator.  Contributing to $\mathcal{L}_\textrm{loss}$ are the collapse operators \cite{carm1999} describing: the spontaneous emissions from state $\ket{3}$ to states $\ket{1}$ and $\ket{2}$ ($\sqrt{\gamma_{31}}  \sigma_{31}$, $\sqrt{\gamma_{32}} \sigma_{32}$);  spin lattice relaxation between states $\ket{1}$  and $\ket{2}$ ($(N_b+1)\sqrt{\gamma_{21}} \sigma_{12}$, $N_b\sqrt{\gamma_{21}} \sigma_{21}$); and the dephasings for the microwave and optical transitions ($\sqrt{\gamma_{2d}} \sigma_{22}$, $\sqrt{\gamma_{3d}} \sigma_{33}$). Here $N_b= (e^{\hbar\omega/(kT)}-1)^{-1}$ is the mean number of bath quanta at the microwave frequency. For our situation, with a 5\,GHz microwave frequency and a 4.2\,K temperature, $N_b \approx 17$.

The steady state coherence on the $\ket{3}\leftrightarrow\ket{1}$ transition is given by $\rho_{31}(\delta_2,\delta_3)$, and it can be obtained from the steady state solution of Eq. (\ref{eq:master}). Then, the total polarization at a given position $z$ and at the $\ket{3}\leftrightarrow\ket{1}$ frequency will be given by 
\begin{equation}
P (z)= N d_{13} \iint d \delta_2 \,d \delta_3 \, g(\delta_2,\delta_3) \rho_{31}(\delta_2,\delta_3,z) +\text{c.c.}~,
\end{equation}
where $N$ is the density of atoms, $d_{ij}$ is the electric dipole moment of the transition between $\ket{i}$ and $ \ket{j}$, and $g(\delta_2,\delta_3)$ describes the distribution of the microwave $(\delta_2)$ and optical $(\delta_3)$ detunings due to inhomogeneous broadening, normalized so that $\iint d\delta_2 d\delta_3\, g(\delta_2,\delta_3) =1$. In our calculations $g(\delta_2,\delta_3)$ is assumed to be a two-dimensional Gaussian function with standard deviations $\Delta_\mu$ and $\Delta_o$. 

This $P(\vec{r}\,)$ acts as a source term and generates a side-band signal via
\begin{equation} \label{eq:source}
\frac{\partial E_S(z)}{\partial z} = \frac{ i \mu_0 \omega_{31} c}{2n} P(z)~,
\end{equation}   
where $\omega_{31}$ is the angular frequency of the $\ket{3}\leftrightarrow\ket{1}$ transition, $n$ is the refractive index of the sample, and $\mu_0$ and $c$ are the magnetic permeability and the speed of light in the vacuum.
In the optically thin limit this differential equation is trivial to solve and the result can be rearranged to give:
\begin{equation} \label{eq:final}
|E_S(z=L)| =\frac{\alpha_{31}L}{2} \cdot \frac{d_{23}}{d_{13}} \cdot \frac{I}{\pi \Omega_{\xi} } \cdot \mathbf{Re} \left(\frac {\int_0^L e^{ink_\mu z}\, dz} {L}\right) \cdot E_\xi~,
\end{equation} 
where $\alpha_{31}$ is the absorption coefficient for the ${\ket{1}\leftrightarrow\ket{3}}$ transition, $L$ is the length of the sample, $E_\xi$ is the amplitude of the coupling laser beam, and ${I = \sqrt{2\pi} \Delta_o\iint d \delta_2 \,d \delta_3 \, g(\delta_2,0) \rho_{31}(\delta_2,\delta_3,z=0)}$. The first and the second terms on the right hand side of Eq.~(\ref{eq:final}) can be easily obtained from experimental absorption measurements for the different transitions. The fourth term is a phase-matching factor due to the fact that the driving laser has a propagating phase of $e^{ink_{32}z}$, while the side-band signal has a propagation phase of $e^{ink_{31}z}$ (where $k_{ij}$ is the wave vectors for the $\ket{\,i\,}\leftrightarrow\ket{\,j\,}$ transition), and it can be calculated very accurately. The third term including $I$ is calculated numerically by solving the master equation as explained above.

The spontaneous emission rates for the two optical transitions used to model the experiment are $\gamma_{31}=60\ \textrm{s}^{-1}$ and $\gamma_{32}=30\ \textrm{s}^{-1}$. These are calculated from the known 11\,ms excited state lifetime and the branching ratios expected from the spin Hamiltonians.

The optical inhomogeneous linewidth is taken from the optical absorption measurements  ($\Delta_o=2\pi\times1$\,GHz), and the spin inhomogeneous linewidth is taken from EPR results ($\Delta_{\mu}=2\pi\times13$\,MHz).

The dephasing rates and the spin lattice relaxation time are not known for this temperature and magnetic field, so they are allowed to vary and the values that gave the best fit to the data are chosen. These values are $\gamma_{3d}=2.8\times 10^6\ \persec$, $\gamma_{2d}=1.7\times 10^6\ \persec$ and $\gamma_{21}=27.4\ \persec$ (1 ms lifetime).

We also introduce two free parameters $\zeta_\mu$ and $\zeta^{-1}_\xi$ which take into account the propagation losses of $P_\mu$ from the setup input to the microwave cavity and the inverse loss of $P_\xi$ from the photodiode detector to the sample. The fitted values for these loss and inverse loss parameters are $\zeta_\mu=$\,13.1\,dB and $\zeta^{-1}_\xi=$\,-6.4\,dB, which are well within the experimental expectations.

\section{Experimental setup}\label{App:setup}
\setcounter{figure}{0}
\setcounter{equation}{0}

\begin{figure}[h]
\centering
\includegraphics[scale=0.31]{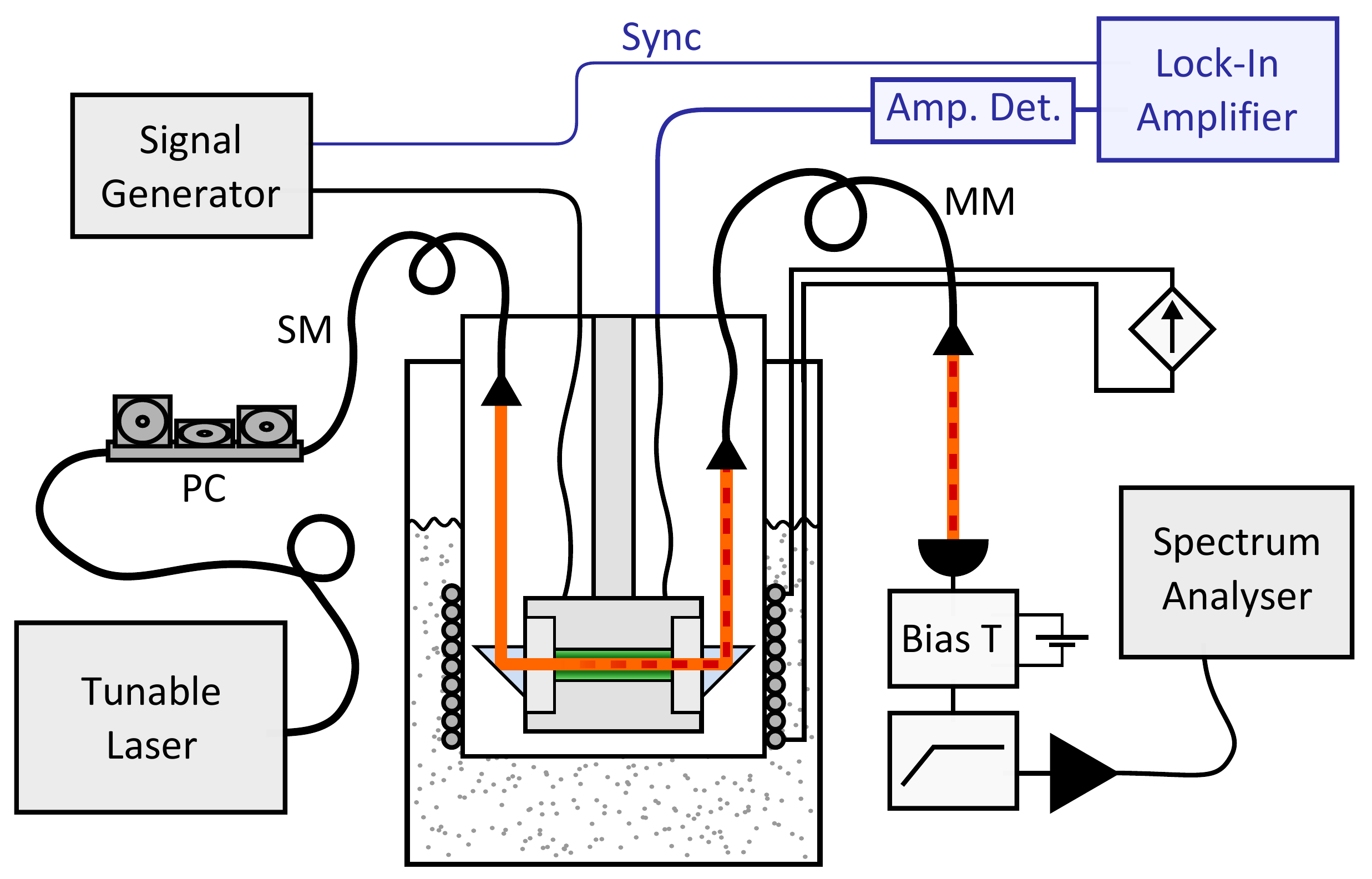}
\caption{\label{fig:setupall} (Color online) Experimental setup. The part drawn in blue is used for EPR measurements only. PC: polarization controller. SM: single mode fibre. MM: multi-mode fibre.}
\end{figure}
The setup for our experiment is depicted in Fig.~\ref{fig:setupall}. The pump beam is generated by a fibre coupled external cavity diode laser at 1536\,nm, and then amplified by an erbium doped fibre amplifier. A polarization controller is used to maximize the output heterodyne signal. The pump beam travels through a single mode fibre and is then collimated using a fibre coupled GRIN lens collimator. With the help of a couple of right angle prisms the beam is sent in and out of the microwave resonator, passing through the Er:YSO sample on its way. The output light, consistent of the pump beam and the up-converted signal is then coupled into a multi-mode fibre using a second GRIN lens collimator. At the output of the multi-mode fibre the light is collimated into a Hamamatsu G7096-03 photodiode detector. A bias tee and a battery serve the double purpose of biasing the photodiode and separating the AC from the DC components of the detected signal. The AC component is then high pass filtered and analysed with a FieldFox N9916A vector network analyser working as a spectrum analyser. To drive the microwave cavity we use an R\&S SMP 22 microwave signal generator. Microwaves are coupled in and out of the cavity using a couple of straight antennas. The cavity itself has a resonant frequency of 4.9\,GHz, a loaded Q factor of around 300 and a room temperature transmission (${\vert S_{21} \vert}^2$) of around 6\,dB.

The resonator and the coupling optics sit inside an encasing stainless steel tube filled with about a mbar of helium, that acts as a thermal exchange gas. This tube is then inserted into a liquid helium bath cryostat. In this way we can avoid optical distortions created by the boiling helium and thermal shocks that could be detrimental to the various optical components. Surrounding the end of the encasing tube is a superconducting magnet powered by a variable current source, which can generate magnetic fields of up to 0.3\,T.

The part drawn in blue in Fig.~\ref{fig:setupall} corresponds to the EPR setup, consisting of an amplitude detector and an SRS SR830 lock-in amplifier. In order to do EPR experiments we add FM modulation to the microwave signal at 3\,kHz. When this FM signal passes through the cavity it gets converted into an AM signal with a modulation amplitude proportional to the slope of the cavity transmission curve. Using and amplitude detector we can measure this modulation amplitude with a lock-in amplifier, which needs to be synchronized with the FM modulation at the signal generator.

The laser source, the signal generator, the lock-in amplifier, the variable current source powering the magnet and the spectrum analyser are all remotely controlled from a computer (not shown in the figure).

\end{document}